\documentclass[a4paper,fleqn,usenatbib]{mnras}

\usepackage{newtxtext,newtxmath}
\usepackage[T1]{fontenc}
\usepackage{ae,aecompl}

\usepackage{graphicx}	% Including figure files
\usepackage{amsmath}	% Advanced maths commands
\usepackage{amssymb}	% Extra maths symbols
\usepackage{pdflscape}

\title[Lunar Occultations at Devashtal]
{Further milliarcsecond resolution results on
cool giants and binary stars from
lunar occultations at Devashtal.
}

\author[A. Richichi et al.]{
A. Richichi,$^{1}$\thanks{E-mail: andrea4work@posteo.eu}
S. Sharma,$^{2}$
T. Sinha,$^{2}$
R. Pandey,$^{2}$
A. Ghosh,$^{2}$
D.K. Ojha,$^{3}$
\newauthor
A.K. Pandey,$^{2}$
and
M.B. Naik$^{3}$
\\
$^{1}$INAF - Osservatorio Astrofisico di Arcetri, Largo E. Fermi 5, 50125 Firenze, Italy \\
$^{2}$Aryabhatta Research Institute of Observational Sciences,
Manora Peak, Naini Tal, 263002 India \\
$^{3}$Tata Institute of Fundamental Research,
Homi Bhabha Road, Colaba, Mumbai - 400005, India
}

\date{Accepted XXX. Received YYY; in original form ZZZ}

\pubyear{2020}

\begin{document}
\label{firstpage}
\pagerange{\pageref{firstpage}--\pageref{lastpage}}
\maketitle

\begin{abstract}
We report on 26 lunar occultation events observed in the
context of a 
program started at Devasthal in 2017. In addition to the
customary observations with the 1.3-m telescope, 
we report here also the first attempts
performed with the TIRCAM2 camera at the
3.6-m DOT telescope in the near-IR. 
The results consist in the first-time angular diameters for 
two late-type giants, in a measurement of the well-known AGB
pulsating variable SW~Vir, and in the measurement
of eight small separation binaries one of which
is detected for the first time (HR~1860). We also measured
the wider binaries SAO~94431 and 55~Tau (twice).
The remaining sources were found to be unresolved
with upper limits as small as 1~milliarcsecond.
We plan
to develop
further the high-speed capability of the TIRCAM2 instrument,
so as to include in our program also more near-infrared, highly extincted sources.
\end{abstract}


%%%%%%%%%%%%%%%%%%%%%%%%%%%%%%%%%%%%%%%%%%%%%%%%%%

\section{Introduction}
The scope and methods of the 
lunar occultation (LO) program at Devasthal have
been described in detail in our previous papers
\citep{2017MNRAS.464..231R, 2018NewA...59...28R}
so that we keep the discussion to a minimum in the
present work.
We use the LO technique mainly for two purposes:
to investigate properties
of late-type giants such as angular diameter, effective
temperature and in the case of very evolved star
the possible presence of circumstellar matter; and to
study sub-arcsecond separation binary stars which are either already 
known but with incomplete data, or discover new systems. In the
case of binary stars, adding observations at epochs, position angles
or filters not hitherto obtained is crucial to complete
the knowledge of the orbit and of the spectral type of the
components.

Although obviously limited in the selection of the sources,
by the one-dimensional projection of the results
and by the fixed-time nature of the events, LO have
several advantages that make them still very attractive.
Among them, the very efficient use of telescope time,
the inexpensive instrumentation, the combination of
high sensitivity and high angular resolution, the relatively simple
and model-independent data reduction, and being largely not
limited by seeing and telescope diameter while achieving
routinely milliarcsecond (mas) resolution.

We report here on 26 LO events, with instrumentation
and data analysis briefly summarised in 
Sect.~\ref{sect:obs_data}.
Eleven of these events include either stars with a resolved angular
diameters or binary stars, which we discuss individually
in Sect.~\ref{sect:results}. We also provide a few
comments on the unresolved sources.

\section{Observations and Data Analysis}\label{sect:obs_data}
All events were recorded at Devasthal (India)
with two different telescopes and instruments,
both operated
by the Aryabhatta Research Institute of Observational Sciences (ARIES).
A log of the observations is provided in 
Table~\ref{tab:observations}, which follows the style 
used in \citet{2018NewA...59...28R}.
The first few columns list the date, time, type of event
(disappearance or reappearance), source designation, V magnitude
and spectrum. These latter were compiled from the {\it Simbad}
database. 

Mainly, we used the 1.3-m telescope 
equipped with 
a  512x512 pixels frame transfer ANDOR iXon 
EMCCD. 
More details were given in \citet{2017MNRAS.464..231R}.
We employed small 
sub-windows positioned at the center
of the detector, with pixel areas as listed under the Sub column
in Table~\ref{tab:observations}.
Rebinning was also employed as listed in column Bin.
The integration time and the effective sampling time in ms
are listed under $\tau$ and $\Delta$T, respectively.
The filters employed were either Cousins R or I.

Also, we employed for the first time for LO work
the near-IR camera TIRCAM2 \citep{2012BASI...40..531N,2018JAI.....750003B}
developed by the
Tata Institute of Fundamental Research (TIFR),
 mounted at the
Cassegrain main port of the 3.6-m ARIES Devasthal Optical Telescope 
\citep[DOT;][]{2018BSRSL..87...29K}.
TIRCAM2 is a closed-cycle Helium cryo-cooled imaging camera 
equipped with a Raytheon 512$\times$512 pixels InSb Aladdin~III
         Quadrant focal plane array (FPA).
         %having sensitivity to photons in the $1-5$\,$\mu$m 
         %wavelength band.
        The field of view (FOV) of the TIRCAM2 is $86\farcs5\times86\farcs5$ 
        with a plate scale of $0\farcs169$.
         The full-width at half-maximum of the stellar images was $\sim 0\farcs7$. 
         TIRCAM2 provides sampling time of $\sim 256$\,ms for the full frame, 
         therefore for LO work we have used a sub-array window of 32$\times$32 pixels
         to increase the readout rate.
The resulting sampling time consists of window readout time ($\sim 6$\,ms) 
as well as other overheads and can vary from 16\,ms to 48\,ms most of the time.
The TIRCAM2 observations, all recorded in a
broad-band K filter,  were obtained at an early
phase of the commissioning of this instrument on the
3.6-m telescope.
In fact, three more observations were attempted 
in October 2017: although the occultations were detected,
the data were not useful for a quantitative analysis
and are therefore not listed in 
Table~\ref{tab:observations}.

%\begin{landscape}
\begin{table*}
	\centering
	\caption{List of observed events}
	\label{tab:observations}
	\begin{tabular}{lcclrlcccrrrl} % 
\hline
Date 	&	Time (UT)	&	Type	&	Source	&	V	&	Sp	&	Filter	&	Sub	&	Bin	&	$\tau$	&  $\Delta$T	&	S/N	&	Notes$^1$ \\
\hline	
2017 Oct 07	&	20:39	&	R	&	$\mu$ Cet	&	4.3	&	A9IIIp	&	R	&	32$\times$32	&	2$\times$2	&	2.28	&	2.75	&	11.4	&	Tm	\\
2017 Oct 08	&	18:21	&	R	&	SAO 93529	&	8.9	&	F8	&	R	&	32$\times$32	&	2$\times$2	&	4.00	&	4.48	&	5.9	&	Bin	\\
2017 Oct 08	&	19:03	&	R	&	SAO 93532	&	6.7	&	G5	&	R	&	32$\times$32	&	2$\times$2	&	3.00	&	3.48	&	38.4	&	UR	\\
2017 Oct 09	&	19:15	&	R	&	SAO 94056	&	8.4	&	N?	&	I	&	32$\times$32	&	2$\times$2	&	4.00	&	4.48	&	30.5	&	UR	\\
2017 Oct 25	&	13:32	&	D	&	IRC -20478	&	4.9	&	K1/2III	&	R	&	16$\times$16	&	2$\times$2	&	2.00	&	2.47	&	67.4	&	Dm	\\
2017 Nov 01	&	12:54	&	D	&	IRC +00008	&	9.7	&	M5	&	I	&	16$\times$16	&	2$\times$2	&	3.00	&	3.47	&	23.1	&	UR	\\
2017 Nov 05	&	21:23	&	R	&	55 Tau	&	6.9	&	F8V	&	R	&	32$\times$32	&	2$\times$2	&	4.00	&	4.48	&	23.3	&	WB	\\
2017 Nov 06	&	00:21	&	R	&	DO 11286	&	10.3	&		&	I	&	32$\times$32	&	2$\times$2	&	4.00	&	4.48	&	9.3	&	Bin	\\
2017 Nov 06	&	17:58	&	R	&	SAO 94431	&	7.2	&	B3V	&	R	&	32$\times$32	&	2$\times$2	&	4.00	&	4.48	&	6.8	&	WB	\\
2017 Dec 01	&	16:35	&	D	&	$\mu$ Cet	&	4.3	&	A9IIIp	&	I	&	16$\times$16	&	2$\times$2	&	2.50	&	2.97	&	52.0	&	Bin	\\
2017 Dec 30	&	14:06	&	D	&	SAO 93803	&	7.2	&	A0	&	R	&	16$\times$16	&	2$\times$2	&	4.00	&	4.47	&	22.9	&	Bin	\\
2017 Dec 30	&	17:14	&	D	&	SAO 93838	&	6.6	&	K1III	&	I	&	16$\times$16	&	2$\times$2	&	3.00	&	3.47	&	41.6	&	UR	\\
2017 Dec 30	&	19:14	&	D	&	55 Tau	&	6.9	&	F8V	&	I	&	16$\times$16	&	2$\times$2	&	4.00	&	4.47	&	22.4	&	WB	\\
2018 Jan 04	&	22:20	&	R	&	$\psi$ Leo	&	5.4	&	M2III	&	I	&	16$\times$16	&	2$\times$2	&	3.00	&	3.48	&	104.7	&	Dm	\\
2018 Jan 08	&	20:49	&	R	&	SW Vir	&	6.9	&	M7III:	&	I	&	32$\times$32	&	2$\times$2	&	2.28	&	2.75	&	26.3	&	Dm	\\
2018 Apr 04	&	19:35	&	R	&	SAO 159887	&	8.9	&	K0III	&	I	&	32$\times$32	&	2$\times$2	&	4.00	&	4.48	&	3.8	&	UR	\\
2018 May 18$^2$	&	14:32	&	D	&	IRC -20156	&	8.5	&	C-N5III:	&	K	&	32$\times$32	&	No	&	6.40	&	16	&	30.3	&	UR	\\
2018 May 21$^2$	&	17:19	&	D	&	SAO 98770	&	9.4	&	F8	&	K	&	32$\times$32	&	No	&	6.40	&	32	&	18.1	&	Bin	\\
2018 Oct 21	&	15:34	&	D	&	HD 221925	&	7.8	&	F0V	&	R	&	32$\times$32	&	2$\times$2	&	3.00	&	3.48	&	5.4	&	Bin	\\
2018 Dec 26	&	19:16	&	R	&	SAO 98568	&	8.0	&	M2.5III	&	R	&	32$\times$32	&	2$\times$2	&	4.00	&	4.48	&	14.8	&	UR	\\
2019 Jan 18	&	16:37	&	D	&	HD 36230	&	8.2	&	K5	&	I	&	16$\times$16	&	2$\times$2	&	3.00	&	3.47	&	36.3	&	UR	\\
2019 Jan 18	&	18:25	&	D	&	HR 1860	&	6.2	&	B6V	&	R	&	16$\times$16	&	2$\times$2	&	2.50	&	2.97	&	19.2	&	Bin	\\
2019 Jan 19	&	16:26	&	D	&	SAO 78514	&	8.0	&	K2	&	I	&	16$\times$16	&	2$\times$2	&	3.00	&	3.48	&	10.3	&	UR	\\
2019 Jan 19	&	18:26	&	D	&	AX Gem	&	9.5	&	M5	&	I	&	16$\times$16	&	2$\times$2	&	3.00	&	3.47	&	5.1	&	Tm	\\
2019 Jan 19	&	19:13	&	D	&	SAO 78587	&	8.6	&	K0	&	I	&	16$\times$16	&	2$\times$2	&	3.50	&	3.97	&	6.7	&	UR	\\
2020 Jan 05	&	15:50	&	D	&	HR 797	&	6.3	&	A2V	&	I	&	21$\times$21	&	2$\times$2	&	3.00	&	3.48	&	17.3	&	Bin	\\

\hline
\end{tabular}
\\

$^{1}$: Tm = event close to the lunar terminator; 
UR = unresolved; Dm = resolved diameter ; Bin = binary ;
WB = wide binary. \\
$^{2}$: TIRCAM2 at DOT 3.6\,m. \\
\end{table*}
%\end{landscape}

The last two columns of 
Table~\ref{tab:observations} list the signal-to-noise
ratio S/N of the best fit, and the type of result.
LO predictions are made with our software using
a wide range of catalogs, and the data enter our archive with
source names which do not follow a particular rule. To 
help in looking them up, we provide in
Table~\ref{tab:crossid} a list of cross-identifications
especially aimed at binary star researchers.
 
%\begin{landscape}
\begin{table*}
	\centering
	\caption{List of source cross-identifications}
	\label{tab:crossid}
	\begin{tabular}{llllll} % 
\hline
$\mu$ Cet	&	HD 17094	&	HIP 12828	&	SAO 110723	&	WDS J02449+1007	&	HR 813	\\
SAO 93529  	&		&	HIP 16991	&		&	WDS J03385+1336AB	&		\\
SAO 93532  	&	HD 22682	&	HIP 17049	&		&		&		\\
SAO 94056  	&	HD 29496	&		&		&		&		\\
IRC -20478  	&	HD 169420	&	HIP 90289	&	SAO 186794	&	WDS J18254-2033AB	&	21 Sgr, HR6896	\\
IRC +00008  	&		&		&	SAO 128708	&		&	FG Psc	\\
55 Tau  	&	HD 27383	&	HIP 20215	&	SAO 93870	&	WDS J04199+1631AB	&		\\
DO 11286  	&		&		&		&		&	TYC 1301-780-1	\\
SAO 94431  	&	HD 34251	&	HIP 24612	&		&	WDS J05167+1826AB	&		\\
%$\mu$ Cet	&	HD 17094	&	HIP 12828	&	SAO 110723	&	WDS J02449+1007	&	HR 813	\\
SAO 93803  	&	HD 26380	&	HIP 19519	&		&		&		\\
SAO 93838  	&	HD 27029	&	HIP 19960	&		&		&		\\
%55 Tau  	&	HD 27383	&	HIP 20215	&	SAO 93870	&	WDS J04199+1631AB	&		\\
$\psi$ Leo	&	HD 84194	&	HIP 47723	&	SAO 98733	&	WDS J09437+1401A	&	16 Leo, HR3866	\\
SW Vir  	&	HD 114961	&	HIP 64569	&	SAO 139236	&		&		\\
SAO 159887  	&	HD 147474	&		&		&		&		\\
IRC -20156  	&	HD 67507	&	HIP 39751	&	SAO 175215	&		&	RU Pup	\\
SAO 98770  	&		&		&		&	WDS J09476+1419AB	&		\\
HD 221925  	&		&	HIP 116488	&	SAO 146801	&	WDS J23363-0707AB	&		\\
SAO 98568  	&	HD 81540	&	HIP 46311	&		&		&		\\
HD 36230  	&		&		&	SAO 77223	&		&		\\
HR 1860  	&	HD 36589	&	HIP 26072	&	SAO 77255	&		&		\\
SAO 78514  	&	HD 46467	&	HIP 31353	&		&		&		\\
AX Gem  	&	HD 260525	&		&		&		&		\\
SAO 78587  	&	HD 261221	&		&		&		&		\\
HR 797  	&	HD 16861	&	HIP 12647	&	SAO 93067	&		&	85 Cet	\\

\hline
\end{tabular}
\end{table*}
%\end{landscape}

In all cases, for our data analysis
we convolved the filter transmission with the response
of the corresponding detector, and we took into account
the finite integration 
time, the primary diameter and the secondary obstruction.

The data products were FITS cubes, which were
converted to photometric light curves using
a digital extraction mask hand-tailored in each
case to the seeing and image motion of the source.
This process was described in 
\citet{2018NewA...59...28R} and references therein.
Only about 1-2\,s of data were analyzed, 
corresponding to an apparent lunar motion of $\approx 2\arcsec$
on the sky.

The light curves were then analyzed using both
a model-dependent (LSM) and a model-independent (CAL)
method. These are recalled e.g. in
\citet{2017MNRAS.464..231R}.
The LSM provides values and errors for free parameters
such as angular diameter, or projected separation and
fluxes for binary and multiple stars. Scintillation
can be accounted for to some extent.
CAL is iterative, and provides a brightness 
profile of the source derived from maximum-likelihood criteria.
The errors associated with the parameters (LSM) and the
uncertainties on the brightness profiles (CAL) have
been discussed in the mentioned references.

\section{Results}\label{sect:results}
The list of stars which we found to be resolved or binary is
provided in chronological order in Table~\ref{tab:results2}, 
which also follows the format used in our previous papers. 
Columns 2 through 6 refer to the geometry of the event,
listing respectively 
the measured rate of the event V,
its deviation from the predicted rate V$_{\rm t}$,
the local lunar limb slope $\psi$, and the effective
position and contact angles, PA and CA respectively. 
Due to limitations in the time sampling,
V could not be computed from the data of the TIRCAM2 observations,
thus the $\psi$ field
is empty and all other quantities are the predicted values.
Column 7 lists the
best-fitting angular diameter
in the uniform disk approximation (UD).
For binaries, columns 8 and 9 list the projected separation
(along the PA direction from primary to secondary) and the brightness ratio, respectively.

In the following subsections we briefly discuss
our results individually for each resolved star,
following the same order as Table~\ref{tab:results2}.
A brief discussion of the unresolved sources is provided
in Sect.~\ref{unresolved}.

%\begin{landscape}
\begin{table*}
	\centering
	\caption{Summary of results: angular sizes (top) and binaries (bottom)\label{tab:results2}}
	\begin{tabular}{lcrrrrrcc} %

%\tablewidth{0pt}

\hline
\multicolumn{1}{c}{(1)}&
\multicolumn{1}{c}{(2)}&
\multicolumn{1}{c}{(3)}&
\multicolumn{1}{c}{(4)}&
\multicolumn{1}{c}{(5)}&
\multicolumn{1}{c}{(6)}&
\multicolumn{1}{c}{(7)}&
\multicolumn{1}{c}{(8)}&
\multicolumn{1}{c}{(9)}\\
\multicolumn{1}{c}{Source}&
\multicolumn{1}{c}{V (m/ms)}&
\multicolumn{1}{c}{(V/V$_{\rm{t}}$)--1}&
\multicolumn{1}{c}{$\psi $($\degr$)}&
\multicolumn{1}{c}{PA($\degr$)}&
\multicolumn{1}{c}{CA($\degr$)}&
\multicolumn{1}{c}{$\phi_{\rm UD}$ (mas)} &
\multicolumn{1}{c}{Proj.Sep.(mas)} & 
\multicolumn{1}{c}{Br. Ratio} \\
\hline
IRC -20478	&	0.4368	&	13.7\%	&	6.0	&	144	&	60	&	2.57$\pm$0.04	&		&		\\
$\psi$ Leo	&	$-0.6282$	&	8.1\%	&	7.3	&	266	&	152	&	3.03$\pm$0.03	&		&		\\
SW Vir$^1$	&	$-0.588$	&	$-$	&	$-$	&	160	&	224	&	\multicolumn{1}{c}{$\approx 17$}	&		&		\\
\hline																	
SAO 93529	&	$-0.7286$	&	2.0\%	&	2.3	&	41	&	154	&		&	20.3$\pm$2.2	&	7.9$\pm$1.4	\\
SAO 94431$^1$	&	$-0.746$	&	$-$	&	$-$	&	51	&	156	&		&	332.6$\pm$1.0	&	3.45$\pm$0.13	\\
55 Tau	&	$-0.5839$	&	$-2.7$\%	&	2.4	&	291	&	214	&		&	137.2$\pm$0.9	&	3.10$\pm$0.05	\\
DO 11286	&	$-0.4578$	&	7.5\%	&	2.7	&	34	&	125	&		&	36.0$\pm1.8$	&	9.33$\pm$0.68	\\
$\mu$ Cet	&	0.5137	&	61.5\%	&	20.2	&	146	&	83	&		&	108.6$\pm$0.3	&	17.9$\pm$0.2	\\
SAO 93803	&	0.5045	&	6.9\%	&	3.1	&	199	&	$-49$	&		&	63.5$\pm$1.3	&	18.1$\pm$1.8	\\
55 Tau	&	0.6569	&	6.1\%	&	4.7	&	47	&	$-33$	&		&	556.3$\pm$3.2	&	2.64$\pm$0.02	\\
SAO 98770$^1$	&	0.806	&	$-$	&	$-$	&	79	&	$-33$	&		&	177.0$\pm$0.4	&	1.29$\pm$0.01	\\
HD 221925$^1$	&	0.341	&	$-$	&	$-$	&	181	&	$-57$	&		&	165.0$\pm$1.4	&	2.61$\pm$0.07	\\
HR 1860$^2$	&	0.4553	&	10.0\%	&	$-5.5$	&	215	&	$-51$	&		&	6.3$\pm$0.3	&	4.63$\pm$0.16	\\
HR 797	&	0.3634	&	10.5\%	&	3.7	&	123	&	63	&		&	11.2$\pm$1.4	&	1.51$\pm$0.14	\\

\hline
\end{tabular}
\\
$^{1}$: V could not be computed from data, using predicted value. 
$^{2}$: possible triple\\
\end{table*}
%\end{landscape}

\subsection{IRC~-20478}\label{irc-20478}
\begin{figure}
	\includegraphics[angle=-90, width=\columnwidth]{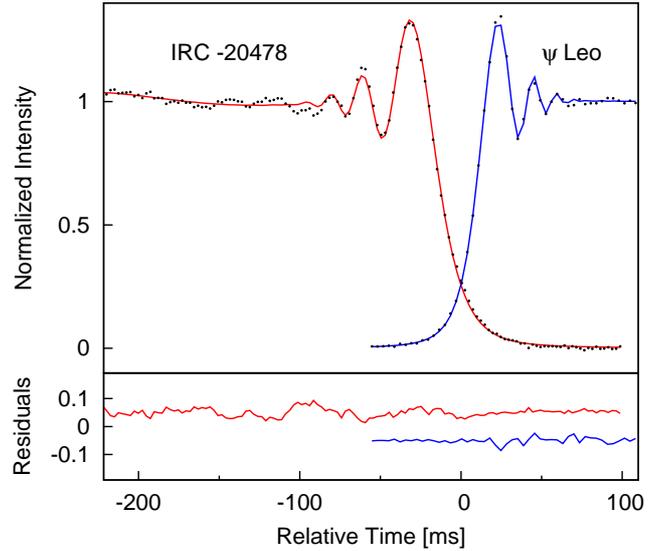}
    \caption{Top panel: left, disappearance
    light curve (dots) for IRC~-20478
    and best fit by a UD model of 2.57 mas (solid line),
    including a scintillation correction by Legendre polynomials. Right,
    reappearance light curve (dots) for $\psi$~Leo and
    best fit by a UD model of 3.03 mas (solid line).
    The two curves have been normalized in intensity, and rescaled
    in x-axis to the time of geometrical occultation. Note that
    the absolute limb speed is about 50\% faster for $\psi$~Leo, 
    see Table~\ref{sect:results}.
    Bottom panel: the fit residuals for the two cases, also
    normalized and offset by arbitrary amounts for clarity.
}
    \label{fig:figure1}
\end{figure}

Only 2 previous LO events were reported for this bright K1/K2 giant, 
without being resolved \citep{1976AJ.....81..650A, 1979ApJS...40..475E}.
We recorded the light curve shown in Fig.~\ref{fig:figure1}
under good conditions in R band, allowing us to
measure this source for the first time.
The data are best fitted 
by an UD angular diameter of $2.57\pm0.04$\,mas, which
can be compared to the
expected value of 2.6~mas 
\citep{1999PASP..111.1515V}.
    The normalized $\chi^2$ value of this fit is 1.3,
    while for comparison it is 2.0 for a fit by a point source.

\subsection{$\psi$~Leo}\label{psi_leo}
No high angular resolution measurements by any technique 
have been reported for this M giant, apart from a mid-IR LO
mentioned by
\citet{1999Msngr..95...25S} who only used it as a test.
We could record a LO light curve with high S/N
as shown in Fig.~\ref{fig:figure1}, leading
to an angular diameter of $3.03\pm0.03$\,mas. The empirical
prediction based on the class and color was 3.6\,mas
\citep{1999PASP..111.1515V}, indicating some possible
discrepancy.
The compilation by \citet{2008MNRAS.389..869E} does not
report suspected multiplicity for $\psi$~Leo.
It was also not revealed as a spectroscopic binary in
a survey by \citet{2009A&A...498..627F}.

\subsection{SW~Vir}\label{sw_vir}
This is a bright AGB star, with a spectral type M7III
and a distance of 300\,pc which result in a very large angular diameter.
%It is in fact
%it is also one of the coolest stars accessible to direct measurements.
As a result,
it has been the subject of numerous investigations
with
a wide range of techniques from LO, to speckle and adaptive optics
on large telescopes, to long-baseline interferometry,
 that we do not
attempt to list here. 
In general, the photospheric
angular diameter is found to be around 15-20\,mas, with a significant
scatter due to the wavelength of observation, to intrinsic pulsation,
and to the presence of a circumstellar shell.

Our goal was to measure not only the diameter but to
also provide a model-independent brightness profile of both
photosphere and circumstellar shell in the style of what was
done e.g. for $\alpha$~Tau by \citet{2017MNRAS.464..231R}.
Unfortunately, this was not possible due to
high-frequency fluctuations in the signal
of a not yet understood nature, possibly caused by wind shaking
of the telescope. 
We could only derive a rough estimate
of the angular diameter of $\approx$17\,mas, in line with
expectations.

\subsection{SAO~93529}\label{sao93529}
A previous LO of this star was reported by
\citet{1981AJ.....86.1277E}, who found it to be a
single star. Later on,
SAO~93529 was suspected as binary from Hipparcos
astrometry and was confirmed in a speckle
measurement by
\citet{2002AJ....123.3442H} with
separation $0\farcs37$ along PA$=88\degr$.
It has since been reported
in several publications,
the latest being by
\citet{2017AJ....153..212H} who found a
separation of $0\farcs30$ along PA$=78\degr$
at epoch 2012.1. Given the
definite motion exhibited over the available time span,
our measurement adds to the database for
a possible orbit computation in the future.
\citet{2017AJ....153..212H} also reported 
$\Delta{\rm (880nm)}=2.7$\,mag, to be 
compared with our $\Delta{\rm R}=2.2$\,mag.

\subsection{SAO~94431 and 55~Tau}\label{55tau}
These two wide visual and speckle binaries have
 already been extensively observed and have
published orbital elements, so that our
measurements are just a mere confirmation.
Perhaps of interest are our accurate flux ratio determinations.
55~Tau is a 
a confirmed member of the
Hyades open cluster \citep{1998A&A...331...81P}.

\subsection{DO~11286}\label{do11286}
This late-type star (IRAS 05255+1832) does not have a spectral classification
in the literature, although its $B-V$, $V-K$ colors
and the GAIA distance of 1.2\,kpc are indicative
of a mid-M type giant.
It was first detected as binary by
\citet{1997A&A...322..202R}, who could obtain
two simultaneous observations in October 1995
from two different sites. The non-detection at
one of the sites was used as a constraint to infer
a $0\farcs1$ to $0\farcs3$ actual separation in
a southerly direction. Upon revisiting these
older data, however, we found that a mistake
was made in the sign of the limb slope.
The actual PA of the 1995 measurement was in fact
322$\degr$ and not 106$\degr$ as reported by
\citet{1997A&A...322..202R}.
The $\Delta$K was  2.2\,mag. No other high
angular resolution measurements are available.

Our present measurement confirms the binary nature
of DO~11286,
and with $\Delta$I of 2.4\,mag, comparable 
to the previously measured  $\Delta$K, it follows that
the spectral
types are possibly quite similar for the two components.
The uncertainty in the estimate of the main spectral type
and possible variability prevent us
from a more precise classification of the components.

If we assume that 
the total system mass is about 1~M$_\odot$
and that the projected separation of $0\farcs036$
in 2017
(or 43~au) is close to the actual one,
it would follow that P $\gtrapprox 280$~yr.
If so, then the 
22.1~yr interval between the two LO events
would be sufficiently small to warrant combining
the projected separations. In this case, we
find the actual separation to be 36.3\,mas
along PA=$41\degr$. Additional measurements
possibly by AO or speckle at a large telescope
would be able to confirm this result.

\subsection{$\mu$~Cet}\label{mu.cet}
This interesting stellar system is a known
speckle binary, additionally found to have 
two more components -- detected so far only
in the near-IR by LO. The situation has
been recently summarized by
\citet{2018MNRAS.478.5683D}, who reported
two LO events observed from the Russian
6-m telescope as well as simultaneous
multi-wavelength speckle observations. 
The same events were observed
also by us at Devashtal, the first one in  October 2017 with
the 1.3-m and the 3.6-m telescopes, and the second
 one in December of the same year with the smaller
 telescope only.
 
 Unfortunately, the first one was with a high
 lunar phase and close to the terminator. Both
 telescopes recorded the event, but TIRCAM2 at
 the
3.6-m was just at the first commissioning
of the LO mode and did not allow us to
derive a useful light curve beyond the simple
detection, while the 1.3-m data are affected
by a rapidly varying background due to the
sunlit lunar mountains on the terminator and
have insufficient S/N to resolve
the system.
 
The December 2017 data are instead of sufficient
quality to resolve at least the speckle companion
with certainty. The parameters are listed in
Table~\ref{tab:results2}. Our $\Delta {\rm I}=3.13$\,mag
is in excellent agreement within errorbars with the LO result
by \citet{2018MNRAS.478.5683D} of 3.16\,mag in
a narrow continuum filter at 850\,nm.
The position angle and projected separation could be
combined with those of the mentioned paper, but in
fact we note that in the presence of a high 
local limb slope of 20$\degr$, which is not too
unusual at the given contact angle, the result
would have a significant uncertainty. This is in
any case less relevant given the extensive speckle
coverage of the system on the same
date by \citet{2018MNRAS.478.5683D}.

\subsection{SAO~93803}\label{sao93803}

Like DO~11286, also SAO~93803 has an 
interesting record of positive and negative binary detections.
They are summarized by
\citet{2017AJ....154..215R}, who  resolved 
the binary once convincingly and once
with borderline S/N in two events
in January and October 2016, respectively,
both in an SDSS $r$ filter.

Our measurement in a Cousin R filter also detects
the companion. \citet{2017AJ....154..215R} estimated
that the period could be of order $\approx$50~yr.
On this premise, if we neglect the 1.9~yr interval and
 combine the January 2016 and December 2017 results
 we obtain a true separation of $0\farcs123$ along
 PA$=78\degr$ at epoch $\approx$2017.0. 
 There remains the intriguing aspect
 of the difference in the brightness ratios, namely
 $\Delta r=2.3$\,mag
 and $\Delta \rm{R}=3.1$\,mag, respectively. This can
 be at least partly explained by the fact 
that the transmission of the $r$ and R filters
 is sufficiently different, with the latter being sensitive to
 less blue and much more
red wavelengths. 
We note that
an AO-aided spectrograph could be used to obtain
individual spectra of the components.

\subsection{SAO~98770}\label{sao98770}
This F8 
star has been known as a binary since several decades, but
has a sparse record of observations by visual micrometer
\citep{1980ApJS...44..111H}, by one previous LO
\citep{1982AJ.....87.1571B}, and by two speckle data points
\citep{1996AJ....112.2260M, 2000AJ....119.3084H}.
The separation of the components has been reported between
$0\farcs15$ and $0\farcs435$, with an upper limit of 
$< 0\farcs036$ in one case.
Only the LO authors
reported a magnitude difference, namely 0.35\,mag in a
RG610 filter.
No orbital elements are known,  and 
there is no available parallax for the SAO~98770 system from
either Hipparcos or GAIA.
The fact that the separation appears to vary
significantly
%even over a few years 
as well as the high proper motion of
almost $0\farcs1$/yr,
however, are suggestive of a rather nearby
system with a relatively short period.

Our measurement complements the set of separations, and
especially it adds the first near-IR magnitude difference
between the components. With $\Delta {\rm K}=0.28$\,mag, the
companion seems to be only slightly cooler than the
primary. 
If the primary is a F8 dwarf, the color difference R$-$K=1.18\,mag
suggests a spectral type around K0V for the secondary and
therefore a combined system mass $\approx 2$\,M$_\odot$.
With this as a constraint,
we find it difficult to reconcile the wide range of measured
separations with periods less than 100~yr, if we have to
assume the distance modulus implied by an F8V main component.
All these facts make SAO~98770
a very interesting candidate for further high
angular resolution studies,
e.g. by speckle at
a large telescope.

\subsection{HD~221925}\label{hd221925}
This relatively wide binary has been reported
in about ten publications
by both visual and speckle observers starting
with \citet{1958JO.....41....1C},
and we add here the first LO measurement.
An orbit determination is still lacking.

\subsection{HR~1860}\label{hr1860}
\begin{figure}
	\includegraphics[angle=-90, width=\columnwidth]{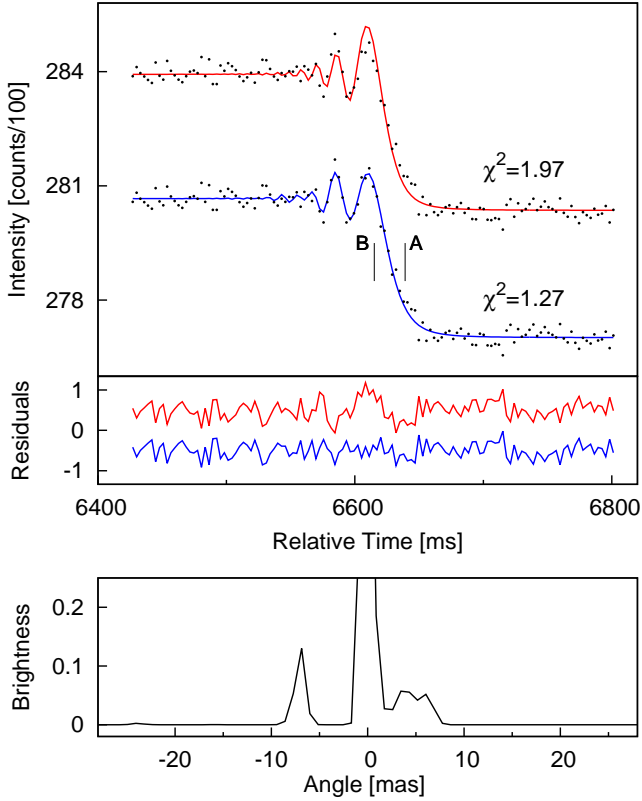}
    \caption{Top panel: light curve (dots) for HR~1860,
    repeated twice with an arbitrary scaling factor and offset. The upper solid line
    is a fit by a point-like source, the lower solid line is 
    the best fit by a binary star, as
    described in the text. The times of occultation of the A, B components
    are marked.
    The normalized $\chi^2$ values for the
    two cases are also shown.
    Middle panel: the residuals for the two fits, offset by
    arbitrary amounts for clarity.
    Bottom panel: brightness profile reconstructed by the CAL 
    method, normalized to 1 at the peak of the primary and
    enlarged for clarity. Note the possible presence of a
    third component, not included in the fit shown.
}
    \label{fig:figure2}
\end{figure}

Our LO data resolve HR~1860 as binary by means of
both model-dependent and model-independent methods, 
as shown in Fig.~\ref{fig:figure2}.
At the GAIA distance of 368\,pc, the projected separation of 6.3\,mas
translates to 2.3\,au. The implication is that the orbital period could
be as short as very few years. Indeed, as shown by the CAL brightness profile
in Fig.~\ref{fig:figure2}, there is an indication of a further component
even closer in. The LSM analysis indicates a fit convergence for
a projected separation of 5\,mas and brightness ratio 1:19 against the
primary, but as the signal is very close to the noise limit and the
improvement in $\chi^2$ is nil, we consider it dubious
and mention the possibility of a triple system only for reference 
to future observers.

No previous high angular resolution measurements are available
for this bright B6 dwarf, with the exception of a LO recorded
in the visual with a 48-cm telescope by \citet{1987SvAL...13..379T}
who found it to be unresolved.
Until now, there had been no indications of binarity 
in this star from other methods such as spectroscopy 
\citep{1984ApJ...285..190A, 2012MNRAS.424.1925C}.
This could be explained by a nearly on-sky inclination
of the system, and/or by a large difference in mass between
the components which would lead to an almost stationary
single-lined spectrum. 

According to the CHARM2 catalog \citep{2005A&A...431..773R}, a total of
116 LO events of 72 unique B stars have been published until 2004, with
31 or 43\% found to be binaries or multiples. This confirms the
high binary fraction already inferred for early, massive spectral types
by \citet{2012MNRAS.424.1925C}.

\subsection{HR~797}\label{hr797}
This source was first detected as a LO binary by
\citet{1982AJ.....87.1571B}, who reported 10\,mas
projected separation along PA=21$\degr$ with
$\Delta{\rm m}=0.18$\,mag in a filter-less blue channel,
from an event in November 1981. Subsequent attempts
by speckle were either partially successful, being
resolved once with 42\,mas separation in V band in 1994
but not in two other observations at similar dates
by \citet{1996AJ....112.2260M}, and not by
\citet{2012AJ....143...42H} in $y$ and I filters.

\begin{figure}
	\includegraphics[angle=-90, width=\columnwidth]{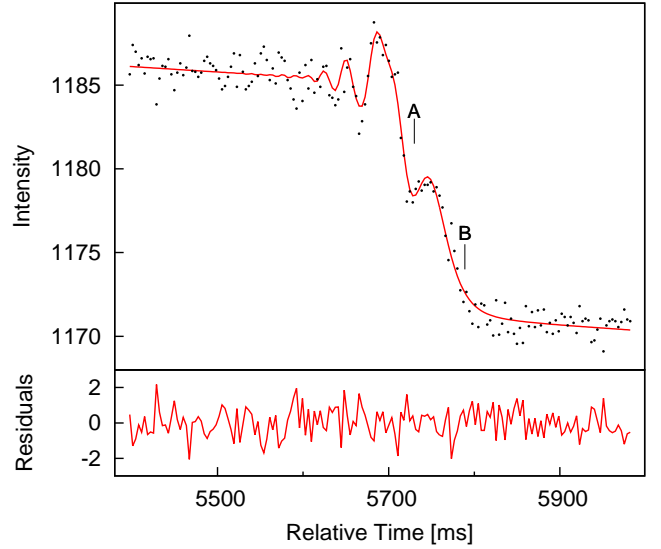}
        \caption{Top panel: light curve (dots) for HR~797,
    and best fit by binary model with 11.6\,mas projected separation.
    The times of occultation of the A, B components
    are marked.
    Bottom panel: the fit residuals.
}
    \label{fig:figure3}
\end{figure}

Our data detection of the binary is
shown in Fig.~\ref{fig:figure3}. At a GAIA distance of
122\,pc, the minimum semi-axis corresponding to the
available (projected and non) separations would
imply a period  shorter than the span between
the few available measurements. Further observations
are needed to determine orbital elements and
dynamical masses. 
We note that this star is included in a
study that uses simulations based on
Hipparcos and GAIA proper motions to derive
constraints on orbits and masses of companions
to nearby stars
\citep{2019A&A...623A..72K}.

\subsection{Unresolved sources}\label{unresolved}
The remaining unresolved sources, some of which were expected
to be binary from previous measurements or resolved in
diameter on the basis of their photometry and spectral type,
are briefly discussed in the following. When applicable,
we compute the upper limit on the angular size
of the source using the algorithm described by
\citet{1996AJ....112.2786R}, and the expected angular
size of late-type giants and evolved stars
according to the empirical formula by
\citet{1999PASP..111.1515V}. 
For this group, we do not list here
all the references in order to be more concise.

SAO~93532 was found unresolved also in two previous LO
events. The same applies to SAO 94056 (two LO events),
SAO~93838, SAO~98568,
HD~36230 and SAO~78587
(one LO event each).
SAO~159887 was a suspected double in one previous
LO event but found unresolved by speckle.
IRC~+00008 and IRC~-20156 have no previous
high angular resolution observations in the literature
and were expected to be resolved with 2.4 and 2.7\,mas,
respectively. The upper limits from our light curves
are 3.8 and 4.5\,mas, respectively.

A previous LO event found  SAO~78514 as binary with
40\,mas projected separation and 1:1.7 brightness
ratio in the 
K band.
The event in the present paper 
shows a ratio $< 10$ in the I band, which
in turn implies a color $\rm{I}-\rm{K} > 3$\,mag
for the secondary. This would indicate a late
M spectral type.
Finally,
AX~Gem was resolved in a previous LO
with a 3.0\,mas UD diameter.
Our present data, obtained at a limb position very
close to the terminator, do not allow us sufficient
S/N to derive a diameter.

We could derive upper limits on the source size
for 9 of the 11 unresolved light curves. The
average upper limit is 3.2\,mas.
For comparison, among the detected binary stars
listed in Table~\ref{tab:results2}, the one with
the smallest upper limit on the angular diameter
of the primary is $\mu$~Cet, with 0.8\,mas.

\section{Conclusions}
We have reported on 26 additional LO measurements carried out
at Devashtal since our latest paper \citep{2018NewA...59...28R}.
In addition to the EMCCD on the 1.3-m telescope,
we have also used for the first time the near-IR camera
TIRCAM2 on the DOT 3.6-m telescope. This instrument  was
in a commissioning phase at the time of our observations
and the time sampling as well as other characteristics
were not optimal yet. Of the five events recorded
with TIRCAM2, we have reported here on two with a
light curve of sufficient quality.

Our results include three resolved angular diameters, two
of which are first time measurements of the late-type
stars IRC-20478 and $\psi$~Leo, and one 
of the well known AGB star SW~Vir. They also include
eight small-separation binaries, of which one is
a new detection (HR~1860). The rest are previously known
but mostly with very few data points and all without
orbital elements, so that our measurements provide
valuable additions especially for what concerns
the flux ratio and color of the companions.
Additionally, we reported on three LO events of
two relatively wide binaries.

Of the remaining 11 unresolved sources, a few
where expected to be resolved either in diameter
or as binaries but we found reasonable justifications
for our negative detections. The mean upper limit
on the angular size of the unresolved sources was
about 3\,mas, and around 1\,mas in the best cases.

The LO method is confirmed as an ideal technique
to achieve high angular resolution comparable to
the most sophisticated facilities such as long-baseline
interferometry, even with a relatively small telescope
and inexpensive commercial instrumentation. The cost
to pay are the obvious limitations in the choice
and repetition of the sources.
We plan to continue our program, and 
to carry out more near-IR LO observations
with an improved fast mode on TIRCAM2 in the near future.

\section*{Data Availability}
The data underlying this article will be shared on reasonable request to the corresponding author.

\section*{Acknowledgments}
This work 
has made use of the SIMBAD database,
operated at CDS, Strasbourg, France.
It is a pleasure
to thank the technical team 
in Devasthal, as well as 
the members of the IR astronomy group at TIFR,
for their support during the observing runs.
DKO and MBN acknowledge the support of the Department of Atomic Energy, 
Government of India, under project No. 12-R\&D-0200.TFR-5.02-0200.

\end{document}